\documentstyle{camera}

\newcommand{\AmS}{{\protect\the\textfont2
  A\kern-.1667em\lower.5ex\hbox{M}\kern-.125emS}}

\newcommand{\beq}{\begin{equation}}
\newcommand{\eeq}{\end{equation}}
\newcommand{\bea}{\begin{eqnarray}}
\newcommand{\eea}{\end{eqnarray}}

\def\dm2{\Delta m^2}
\def\sq2{sin^2(2\Theta)}

\input epsf

\begin{document}

%
\title{Radiative Transfer in Black Hole Systems}

%
\author{KINWAH WU$^1$, WARRICK BALL$^2$ \And STEVEN V. FUERST$^1$}

%
\organization{
1. Mullard Space Science Laboratory, University College London, 
    Holmbury St Mary, Surrey RH5 6NT, UK \\ 
 2. Department of Physics and Astronomy, University College London, 
    Gower Street, London WC1E 6BT, UK \\ 
3. KIPAC, Stanford University, Stanford, CA 94305, USA}

\maketitle

\begin{abstract}
We present a convariant formulation for radiative transfer in curved space time 
 and demonstrate some applications in the black-hole systems. 
We calculate the emission from semi-transparent accretion tori around black holes, 
  for opacity provided by the Fe K$\alpha$ and K$\beta$ lines and 
  for opacity dominated by electron scattering. 
We also calculate the emission from radiative inefficient accretion flow in 
   black holes with opacity provided by electron-positron annihilation lines.  
Finally we show shadows cast by accreting black holes with different spins 
  and with different distribution of warm material around them. 
\end{abstract}
\vspace{1.0cm}

\section{General Relativistic Radiative Transfer} 

A black hole is characterised by a singularity and an event horizon. 
The presence of black holes in this universe is based on inferences, 
  as no black hole has been observed directly. 
Now the question is: what does a black hole look like? 
Perhaps, it may be more meaningful to ask: 
 what does an astrophysical black hole look like if we can image it, and 
 can we see the shadow of its event horizon? 
 
Black holes are relativistic objects. 
A convariant radiative transfer formulation is therefore needed 
  to calculate the radiation from a black hole's vicinity. 
In the absence of scattering, the transfer equation reads 
\begin{equation}
   \frac{1}{k^\delta u_\delta}  
   \left[ k^\alpha \frac{\partial {\cal I}}{\partial x^\alpha} 
       - \Gamma^\alpha_{\beta \gamma} k^\beta k^\gamma 
        \frac{\partial{\cal I}}{\partial k^\alpha}\right]    
          =  -\chi_0 (x^\beta, k^\beta){\cal I}  + \eta_0 (x^\beta, k^\beta)                                                                                                           
\end{equation}  
    (Lindquist 1966; 
    Baschek et al.\ 1997; 
    Fuerst and Wu 2004, 2007; Wu and Fuerst 2008), 
    where $k^\alpha$ is the 4-momentum of the photon 
    and $u^\alpha$ is the 4-velocity of the medium. 
The specific intensity ${\cal I}$, absorption coefficient $\chi_0$ and emission coefficient $\eta_0$
   are in Lorentz invariant forms.   
The left side of the equation describes the radiation propagation;   
  the right side specifies how the radiation interacts with the medium.  
When the velocity fields of the emitters and absorbers are given, 
  the transfer equation can be solved along the geodesics using a ray-tracing algorithm.  

Scattering deflects photons into the line-of-sight. 
In its presence the radiative transfer equation is an integro-differential equation. 
Fuerst (2005) used projection moment tensors 
  to decompose the transfer equation into a set of moment equations. 
The first two moment equations are
\begin{eqnarray} 
  m_\alpha \left[ {{\cal J}^\alpha}_{, \beta} n^\beta 
     +\Gamma^\alpha_{\gamma \beta} {\cal J}^\beta m^\gamma 
        + \xi \left(1 - \frac{\partial}{\partial \ln E} \right){\cal J}^\alpha \right] 
           \hspace*{2.25cm} &  &   \nonumber \\ 
      & & \hspace*{-10.45cm} 
      =  \rho \left[ - \left(\sigma_{\rm x} +\sigma_{\rm sc}\right) {\cal J}^\alpha m_\alpha   
        + \sigma_{\rm x}{\cal S}_{\rm x} 
        + \sigma_{\rm sc} {\cal J}^\alpha u_\alpha   \right] \ ;       \\ 
 m_\alpha m_\beta \left[ {{\cal J}^{\alpha \beta}}_{, \gamma} n^\gamma 
     +2 \Gamma^\alpha_{\gamma \delta } {\cal J}^{\delta \beta} m^\gamma 
        + \xi \left(2 - \frac{\partial}{\partial \ln E} \right){\cal J}^{\alpha \beta} \right] 
           \hspace*{1cm} &  &   \nonumber \\ 
      & & \hspace*{-10.5cm} 
      =  \rho \left[ -\left(\sigma_{\rm x} +\sigma_{\rm sc}\right) {\cal J}^{\alpha \beta} m_\alpha m_\beta   
        + \sigma_{\rm x}{\cal S}_{\rm x} 
        + \sigma_{\rm sc} {\cal J}^{\alpha  \beta} 
         \left(u_\alpha u_\beta +\frac{1}{10} n_\alpha n_\beta  \right) \right]        
\end{eqnarray}   
 respectively, where $\rho$ is the density of the medium, 
    $\sigma_{\rm x}$ is the absorption cross section, 
    $\sigma_{\rm sc}$ is the electron scattering cross section, 
    ${\cal S}_{\rm x}$ is the source function, 
    $n^\alpha$ is the photon directional unit vector ($n^\alpha n_\alpha = 1$), 
    $E = -k^\alpha u_\alpha$,  
    $m^\alpha = n^\alpha + u^\alpha = k^\alpha/E$,  
    and $\xi = {u^\alpha n_\alpha}_{; \beta} n^\beta 
    + {n^\alpha u_\alpha}_{; \beta} u^\beta$.   
(Here, we assume that electron scattering is the prime scattering process.)
The j-th order specific intensity is 
\begin{equation} 
   {\cal I}_{\rm j }(x^\beta, k^\beta)\ = 
     {\cal J}_{\alpha_1 \alpha_2 .... \alpha_{\rm j}}~  
        m^{\alpha_1} m^{\alpha_2} ...... m^{\alpha_{\rm j}}      
\end{equation} 
  (Fuerst 2005; Wu et al.\ 2006).  
The moment equations are not closed. 
However, they would converge rapidly for certain choices of moment functions.  
In practical calculations, 
  using a truncated set of low order moments is usually sufficient. 
  
\begin{figure}[ht!]  
\begin{center}
\epsfxsize=3.6cm \hspace{0.00cm} \epsfbox{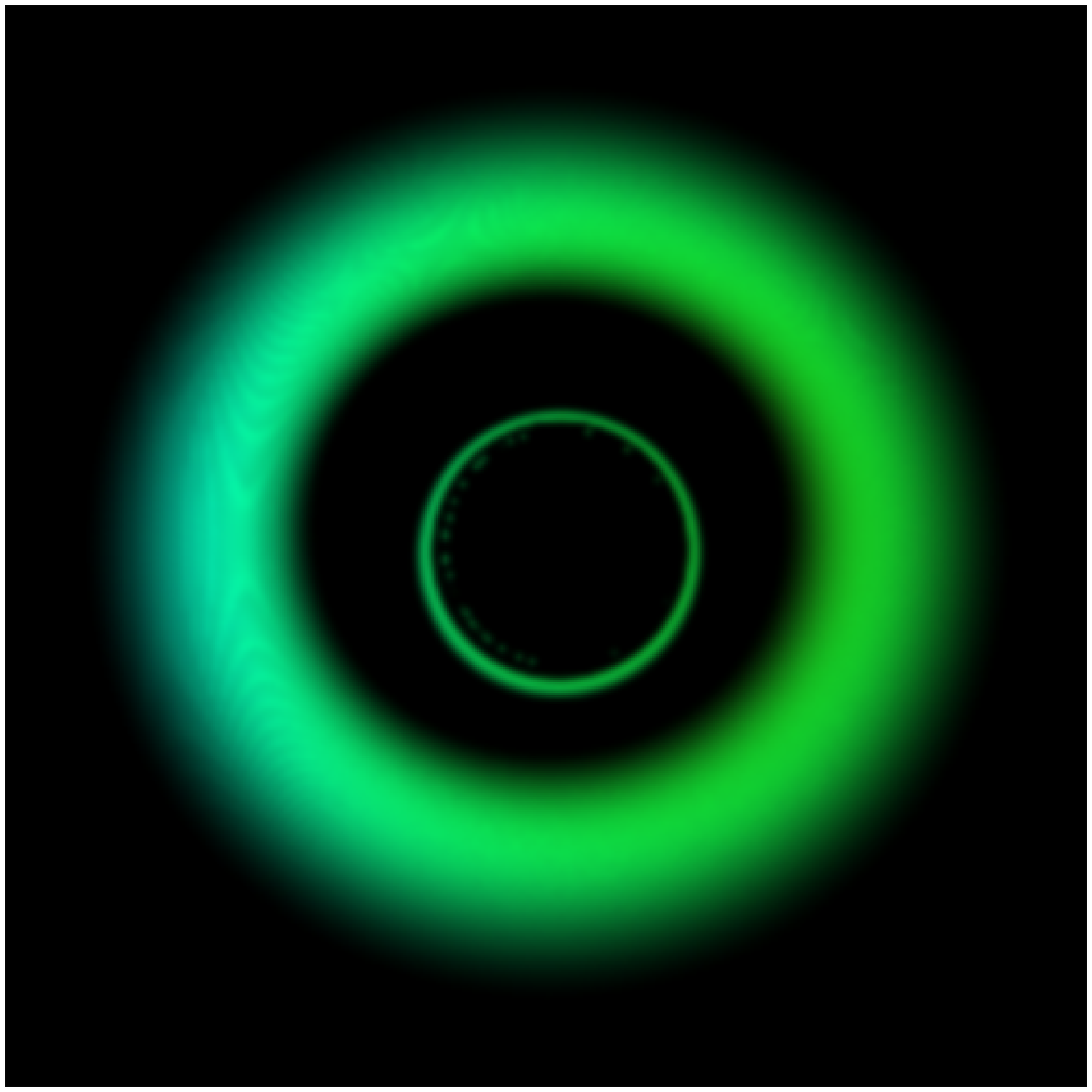} 
\epsfxsize=3.6cm \hspace{0.0cm} \epsfbox{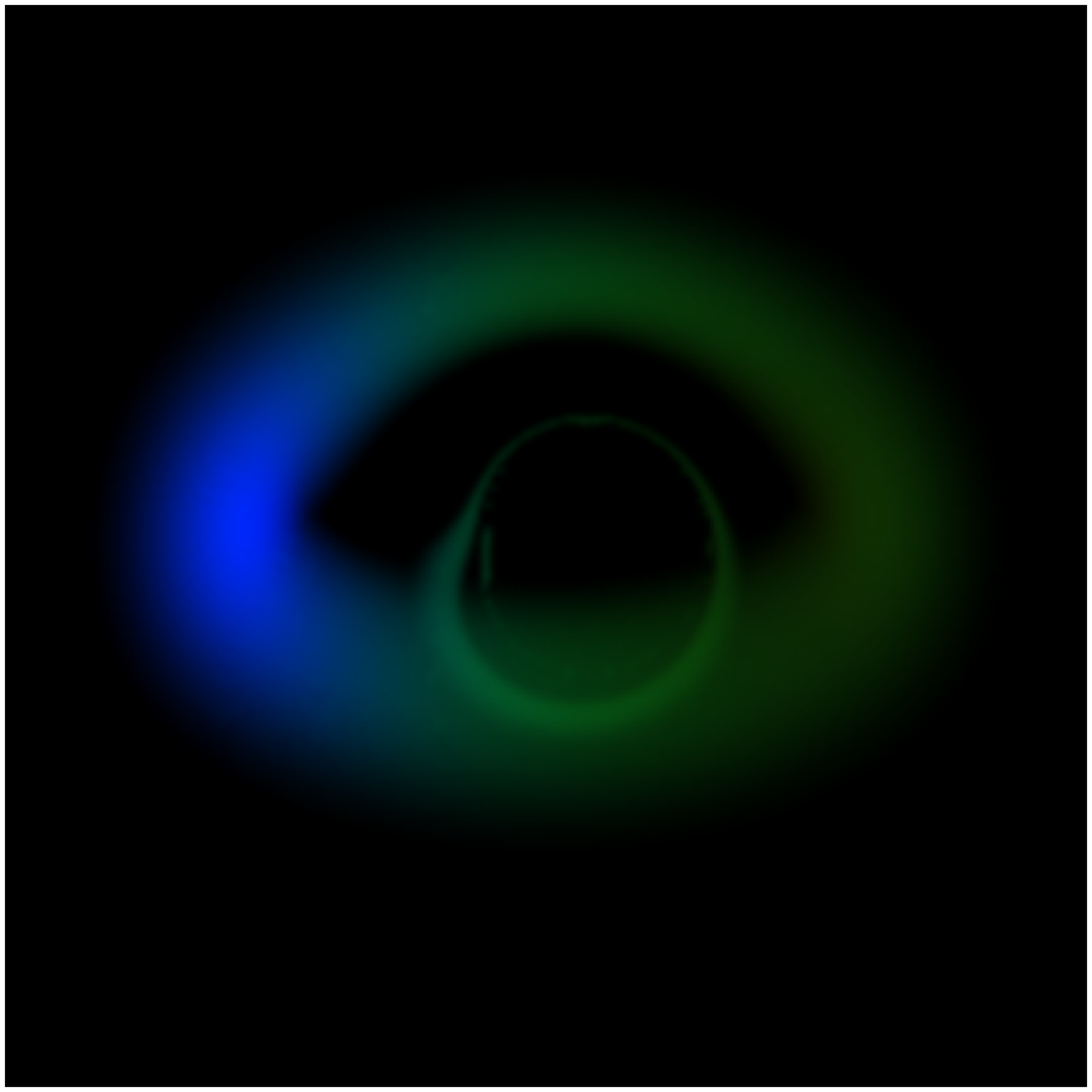} 
\epsfxsize=3.6cm \hspace{0.0cm} \epsfbox{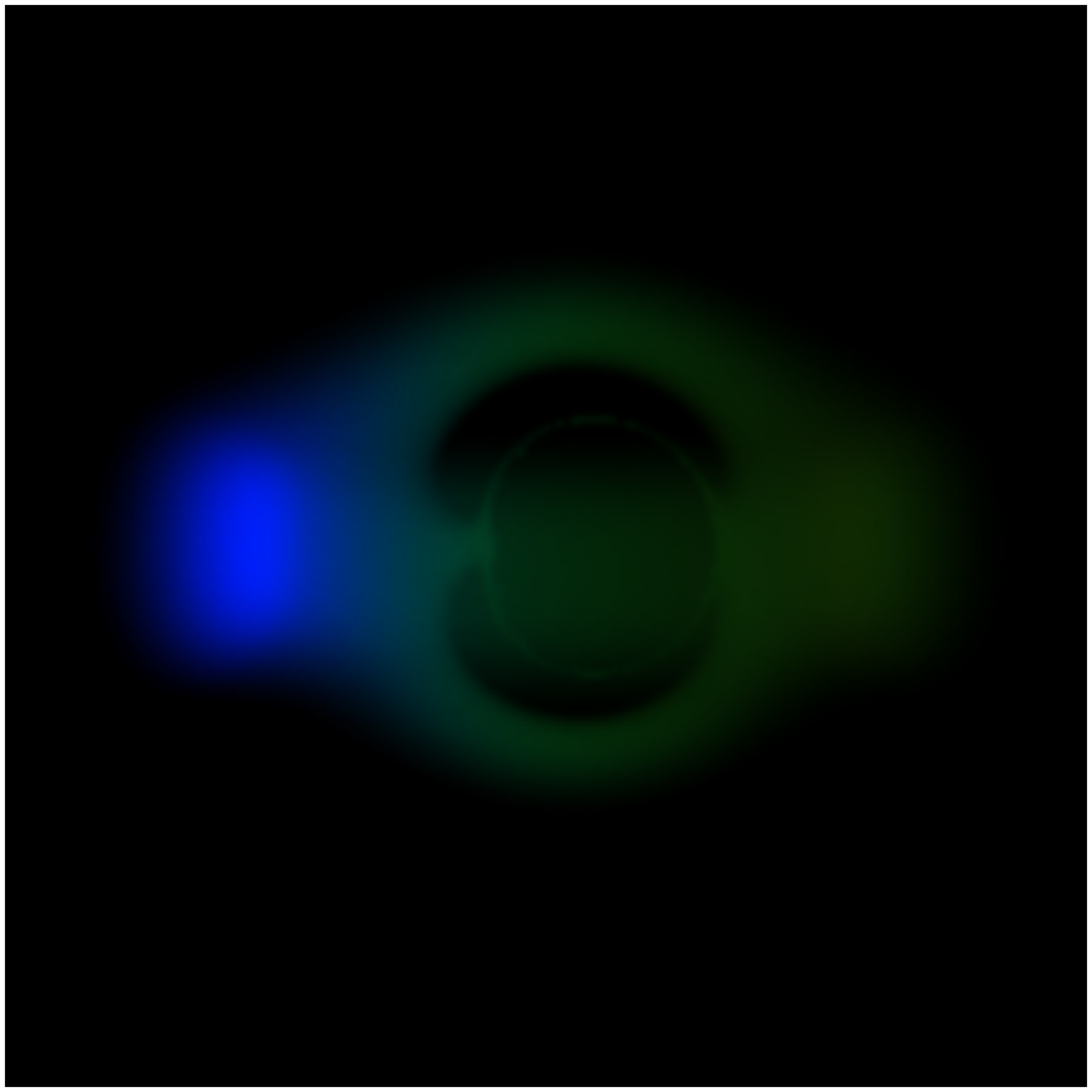} 
\end{center}
\vspace*{0cm} 
\caption[h]{Surface brightness of a semi-transparent accretion torus 
       around a Kerr black hole ($a=0.998$), viewed at inclinations 
       of $15^\circ$, $60^\circ$ and $85^\circ$ (left to right). 
    The opacity is due to the Fe K$\alpha$ and K$\beta$ lines. 
    The brightness is normalised, with all tori 
        having equal maximum pixel brightness.      }
\end{figure}

\begin{figure}[ht!]  
\begin{center}
\epsfxsize=3.6cm \hspace{0.00cm} \epsfbox{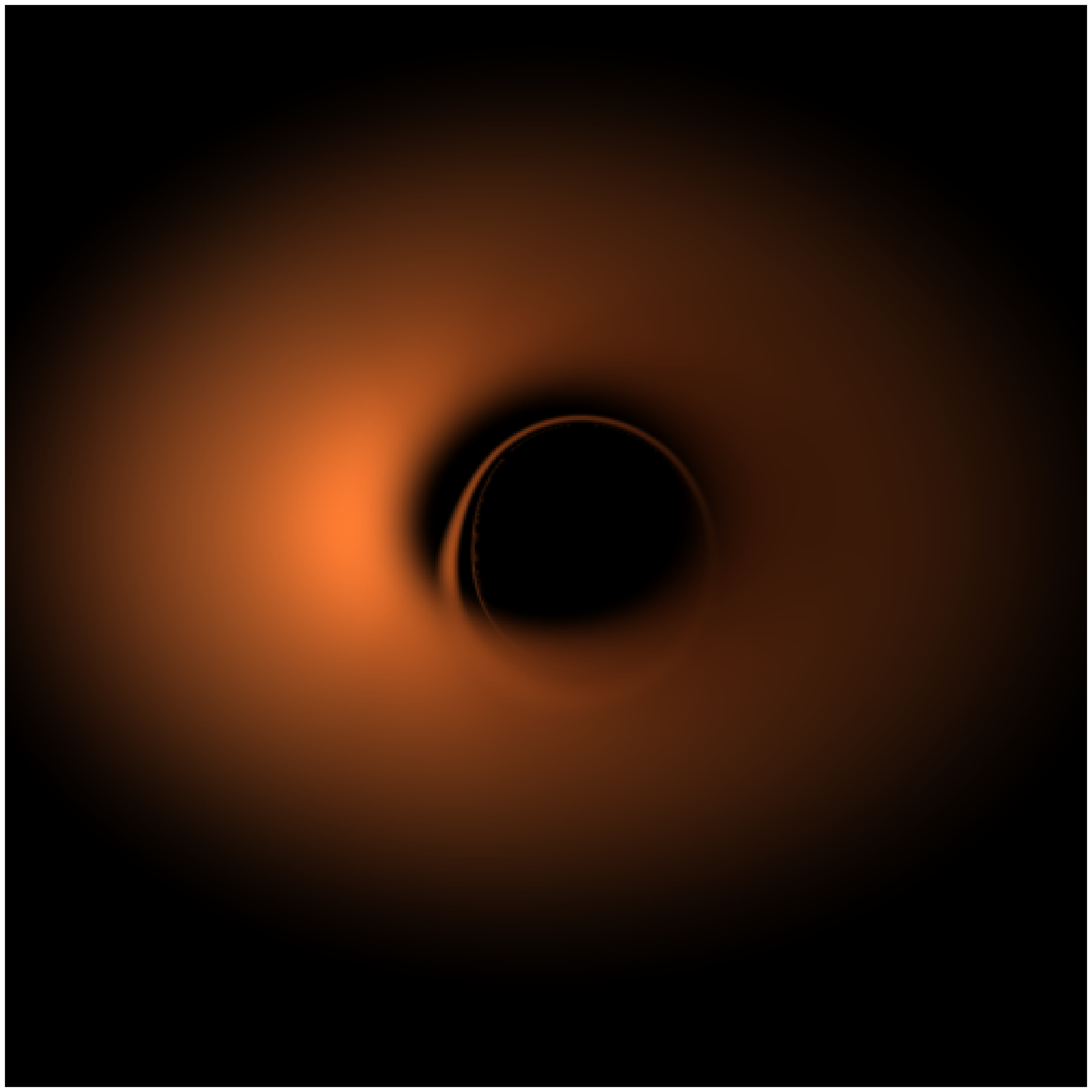} 
\epsfxsize=3.6cm \hspace{0.0cm} \epsfbox{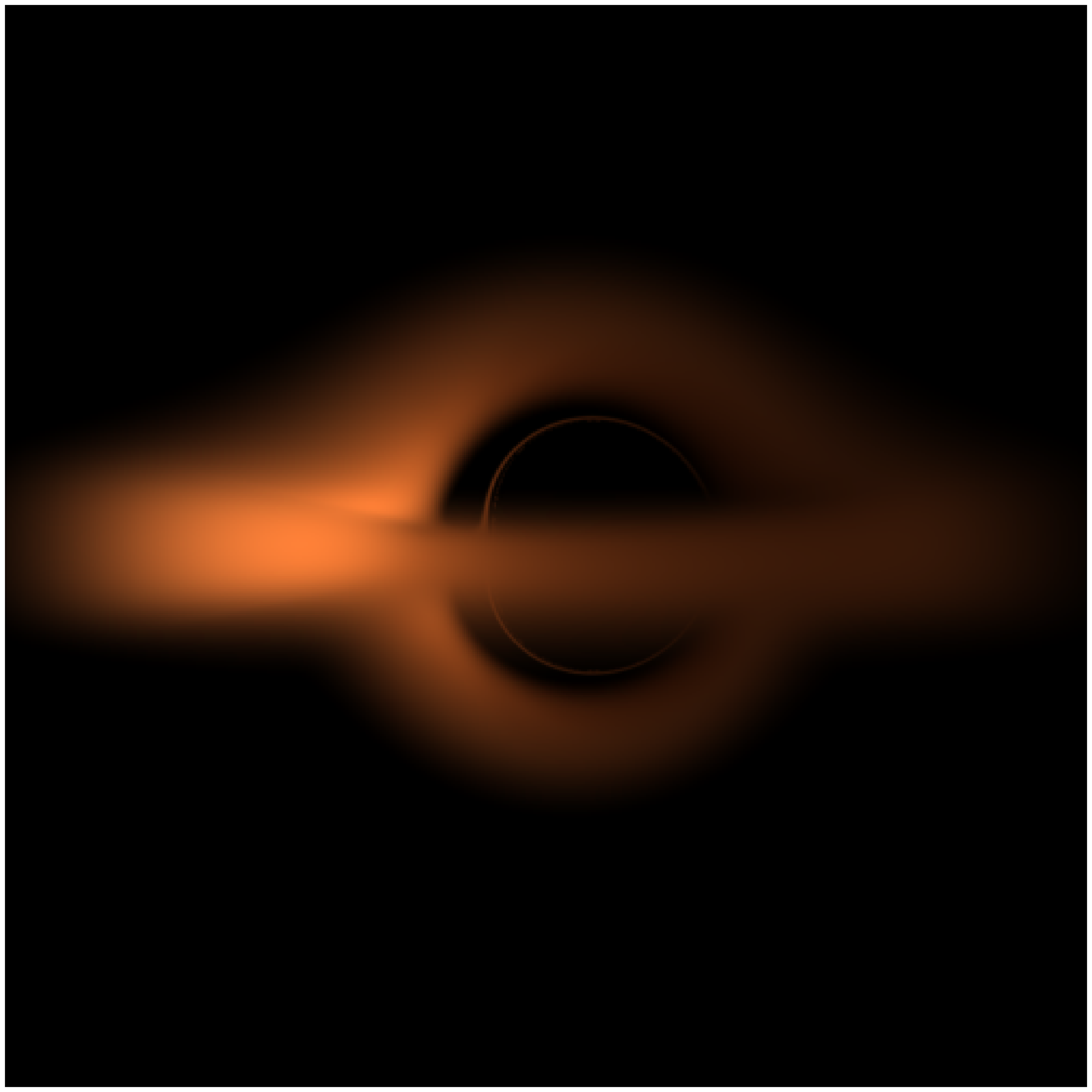}  
\end{center}
\vspace*{0cm} 
\caption[h]{Surface brightness of a semi-transparent electron scattering 
       accretion torus around a Kerr black hole ($a=0.998$), 
       viewed at inclinations of $45^\circ$ (left) and $85^\circ$ (right). 
   Brightness  normalisation is the same as that in Fig.~1. }
\end{figure} 

\begin{figure}[ht!]  
\begin{center}
\epsfxsize=3.75cm \hspace{0.0cm} \epsfbox{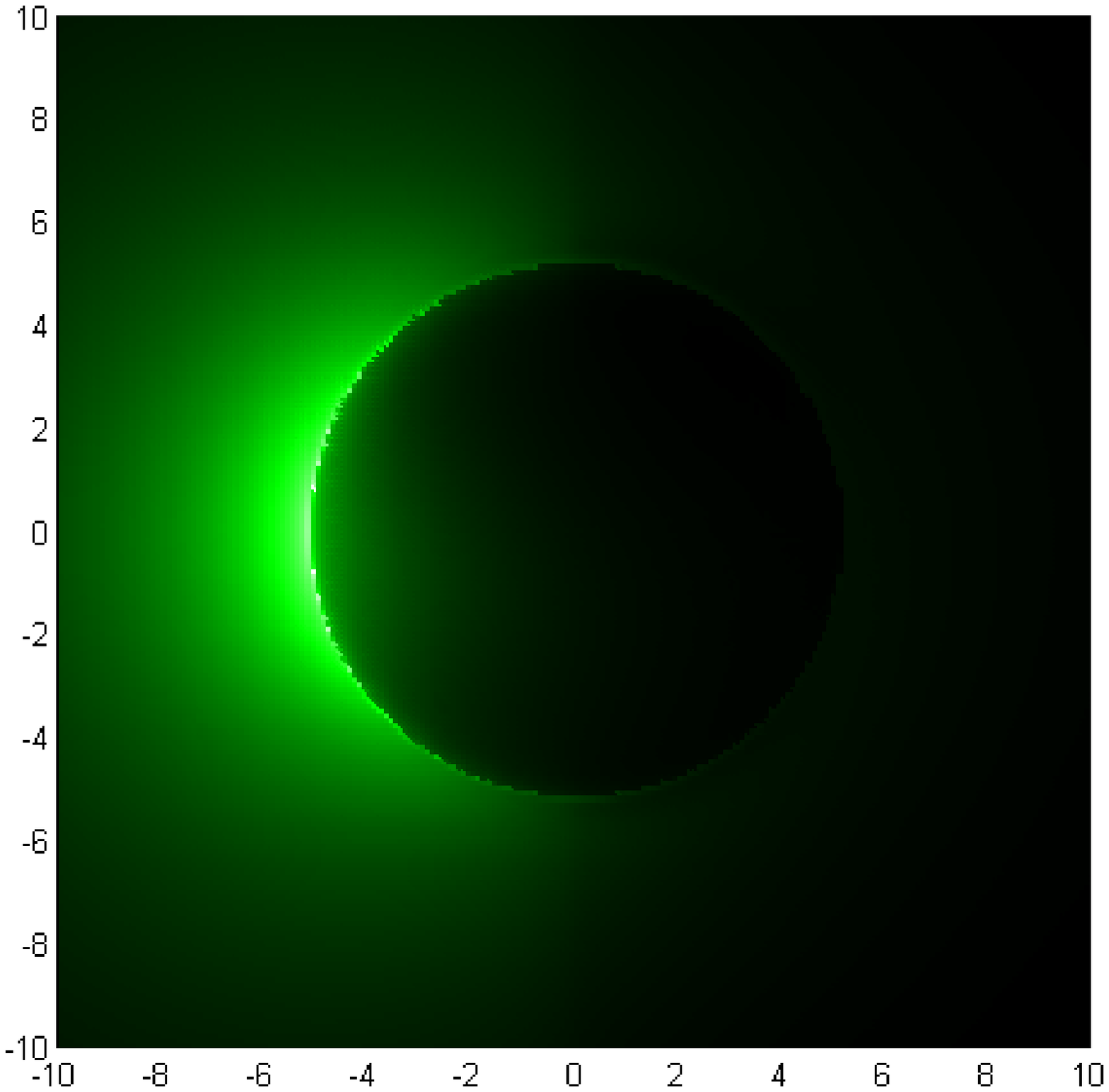} 
\epsfxsize=3.75cm \hspace{0.0cm} \epsfbox{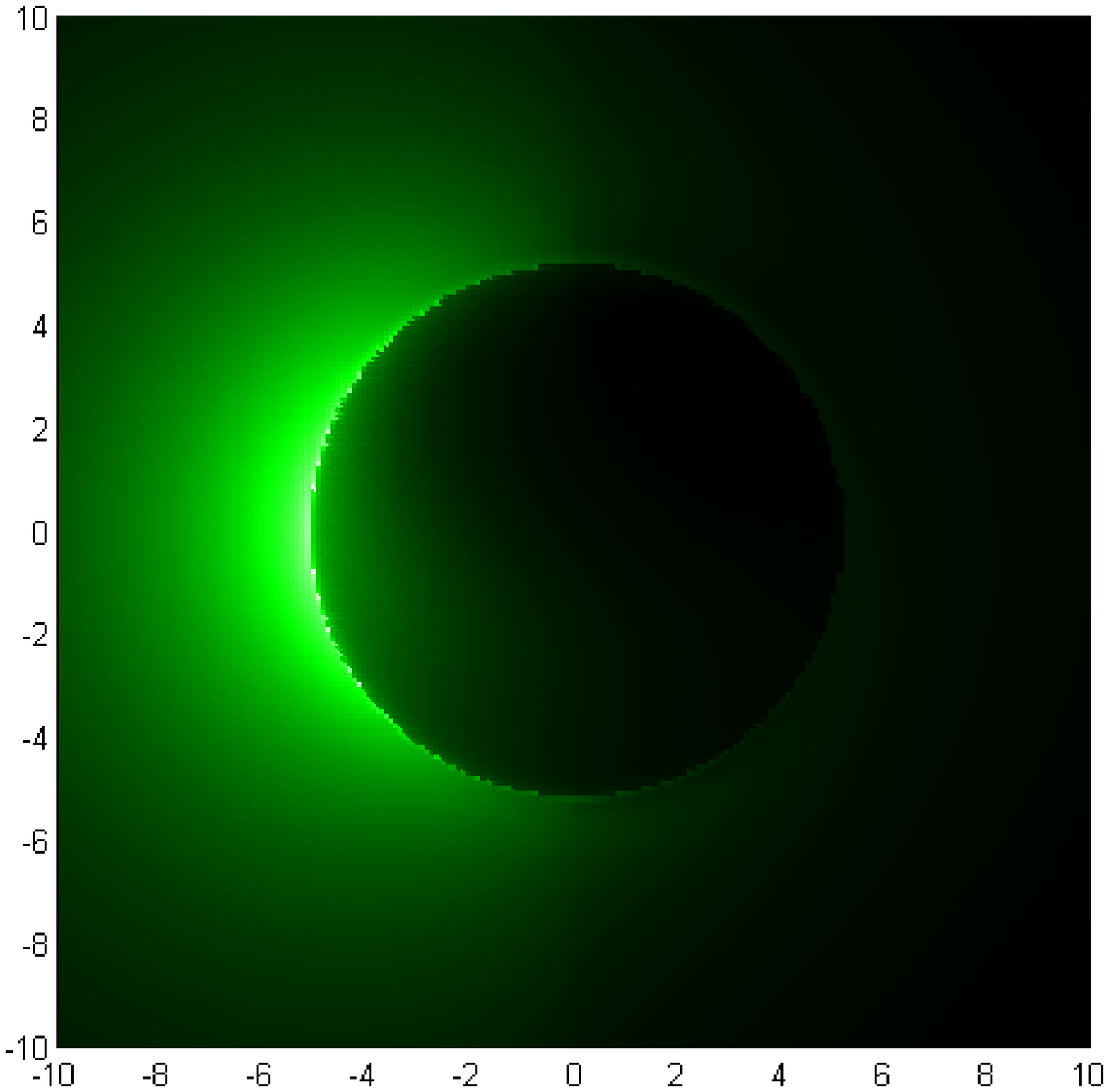}  
\epsfxsize=3.75cm \hspace{0.0cm} \epsfbox{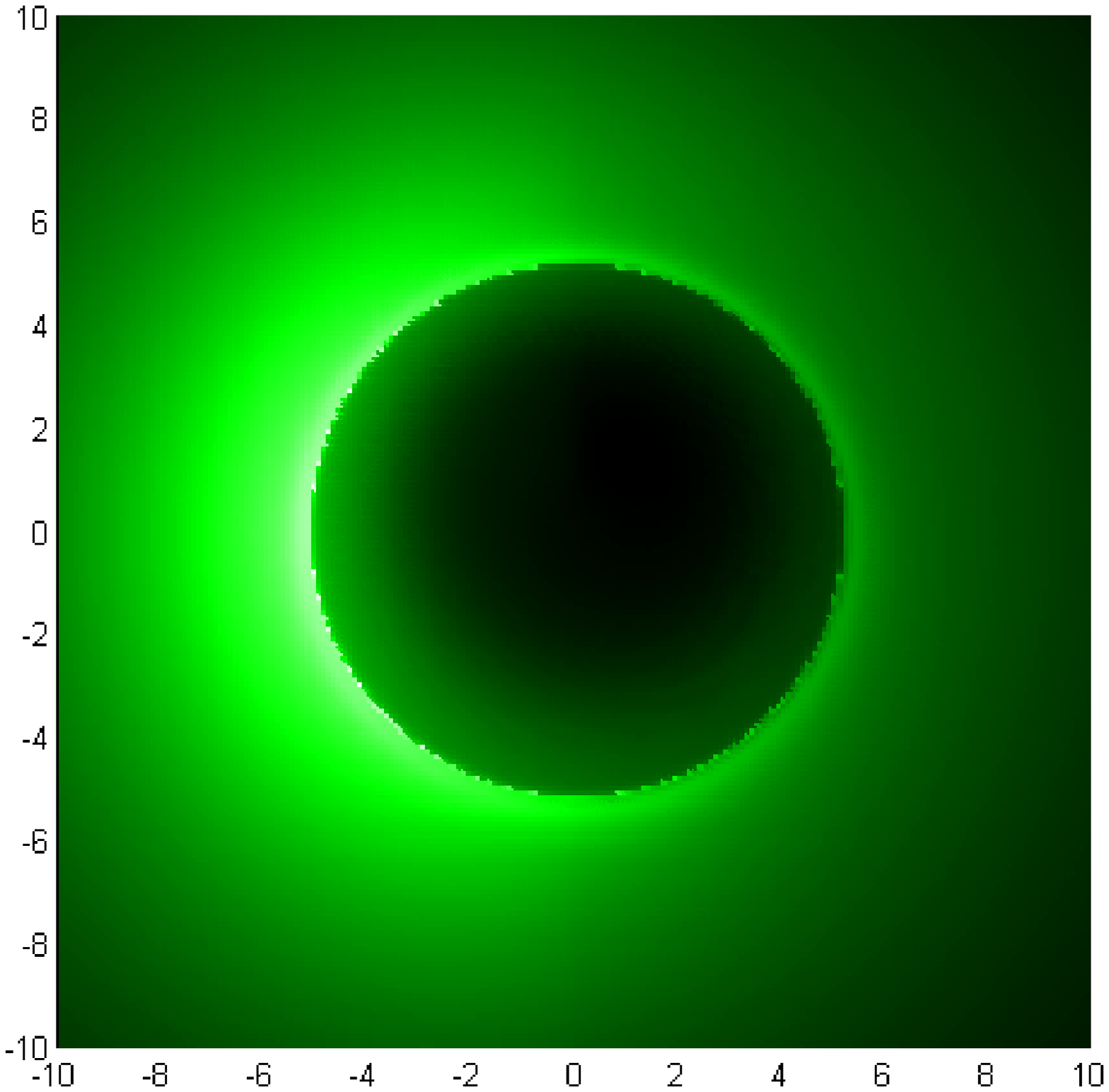} \\ 
\epsfxsize=3.75cm \hspace{0.0cm} \epsfbox{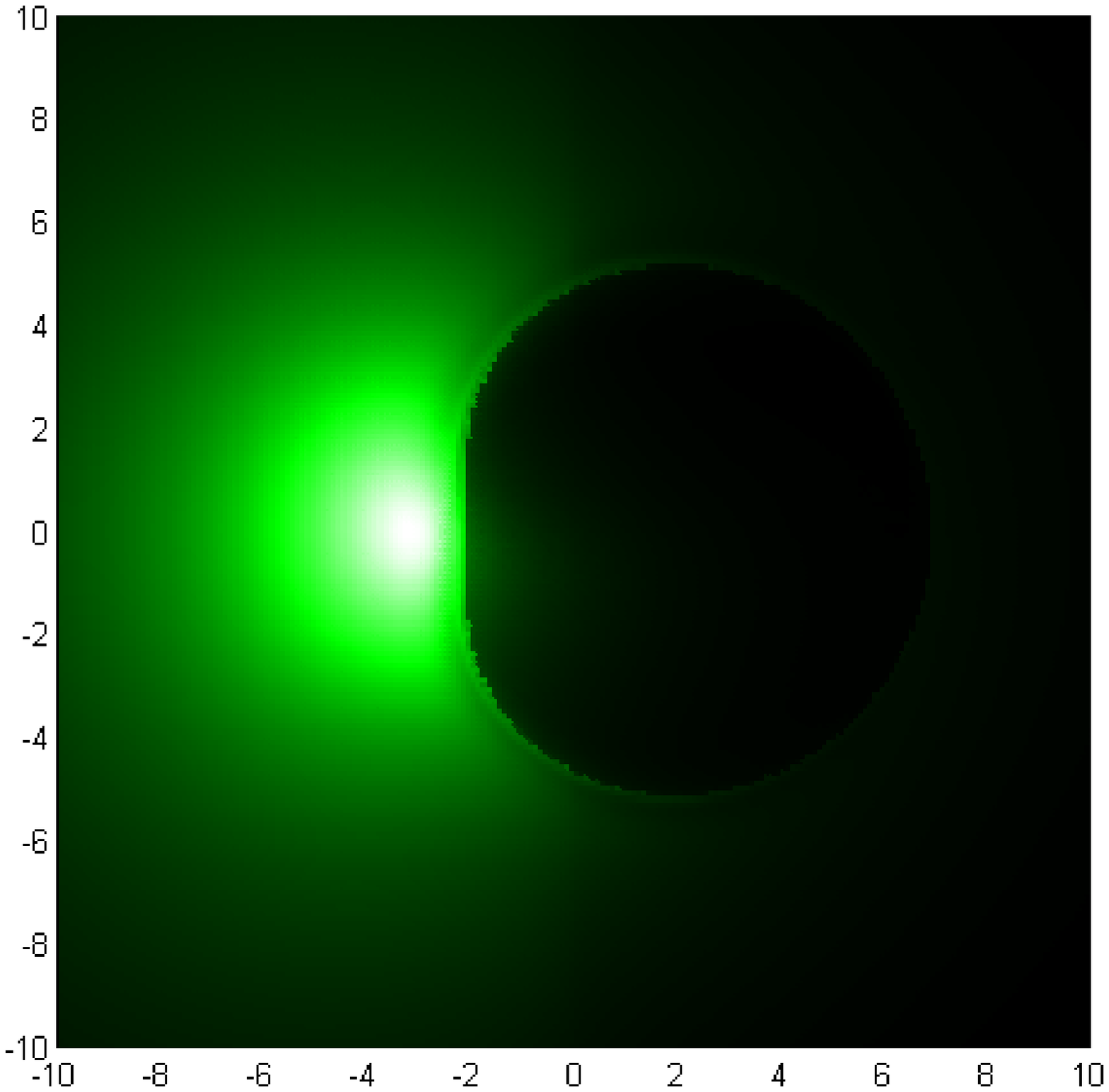}  
\epsfxsize=3.75cm \hspace{0.0cm} \epsfbox{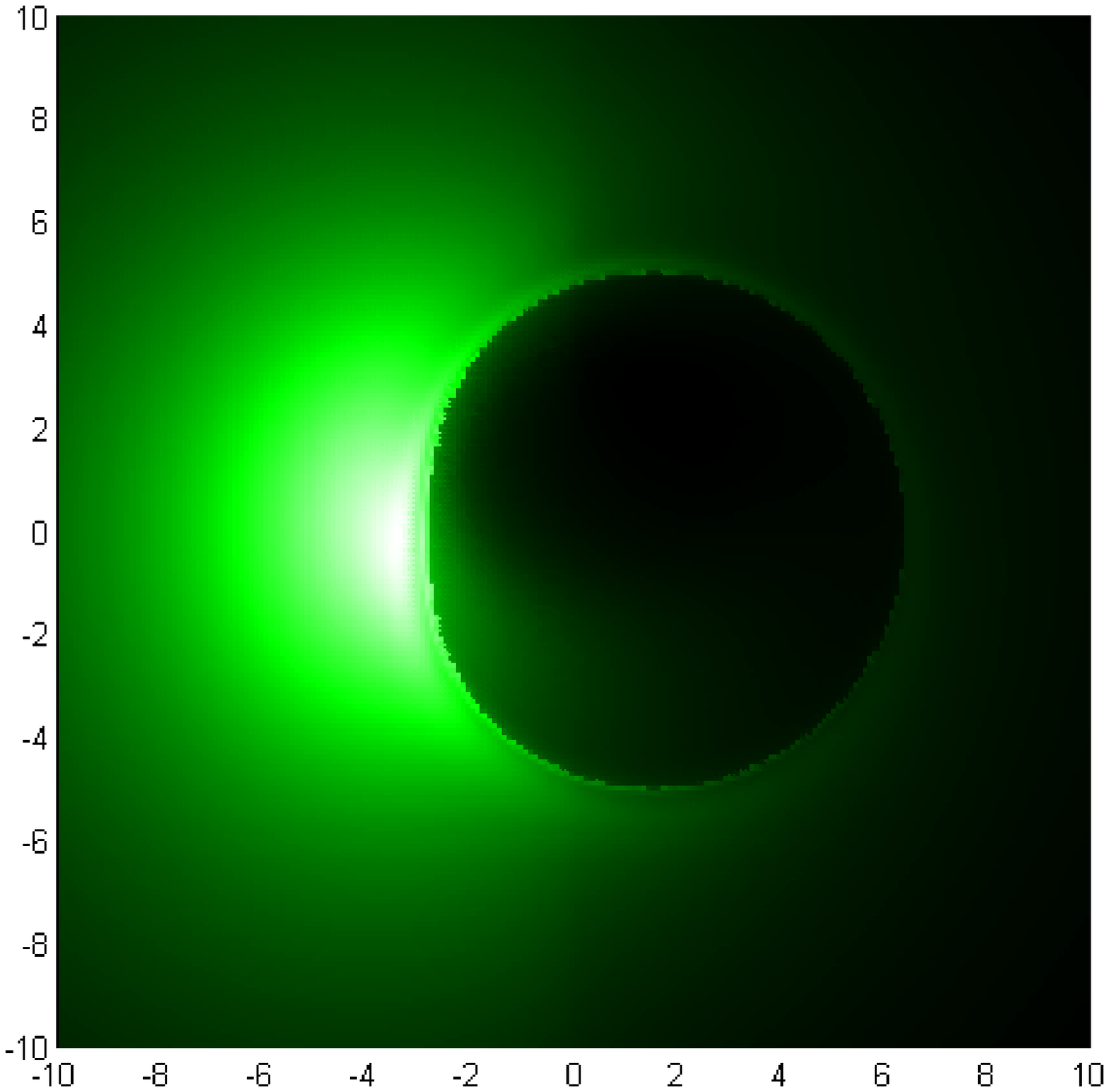} 
\epsfxsize=3.75cm \hspace{0.0cm} \epsfbox{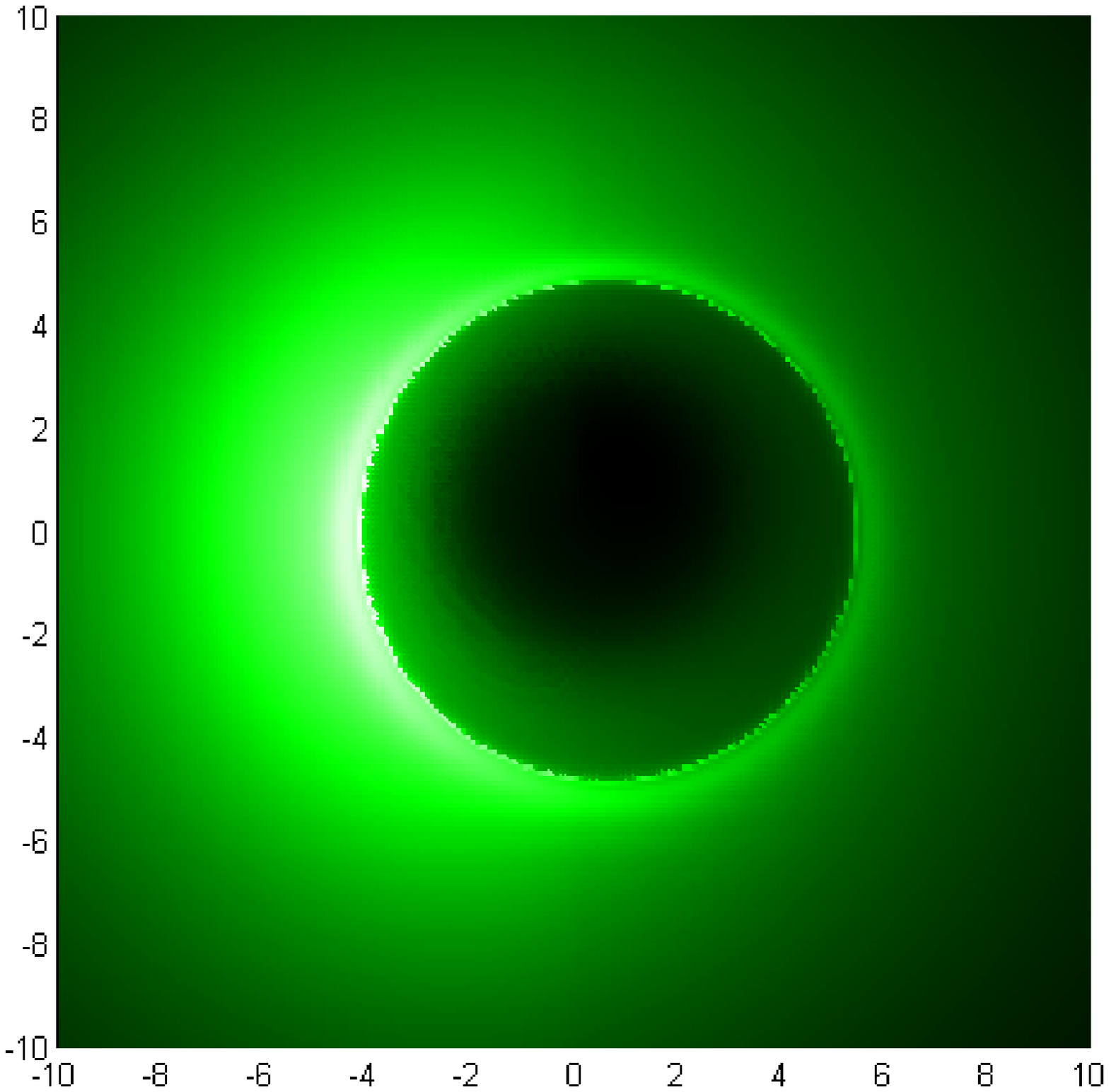}  
\end{center}
\vspace*{0cm} 
\caption[h]{Intensity images of emission lines from vertically Gaussian-distributed 
  hot plasmas around a Schwarzschild black hole (top row) and a Kerr black hole 
    with $a = 0.998$ (bottom row). 
  The density distribution takes the form: $n(r,z)  \propto z^{-3/2} \exp(-z^2/H^2)$.  
       The viewing inclinations are $75^\circ$, $45^\circ$ and $15^\circ$ 
       (left to right columns). 
    The opacity is provided by the electron-positron annihilation line.   }
\end{figure}

\begin{figure}[ht!]  
\begin{center}
\epsfxsize=3.75cm \hspace{0.0cm} \epsfbox{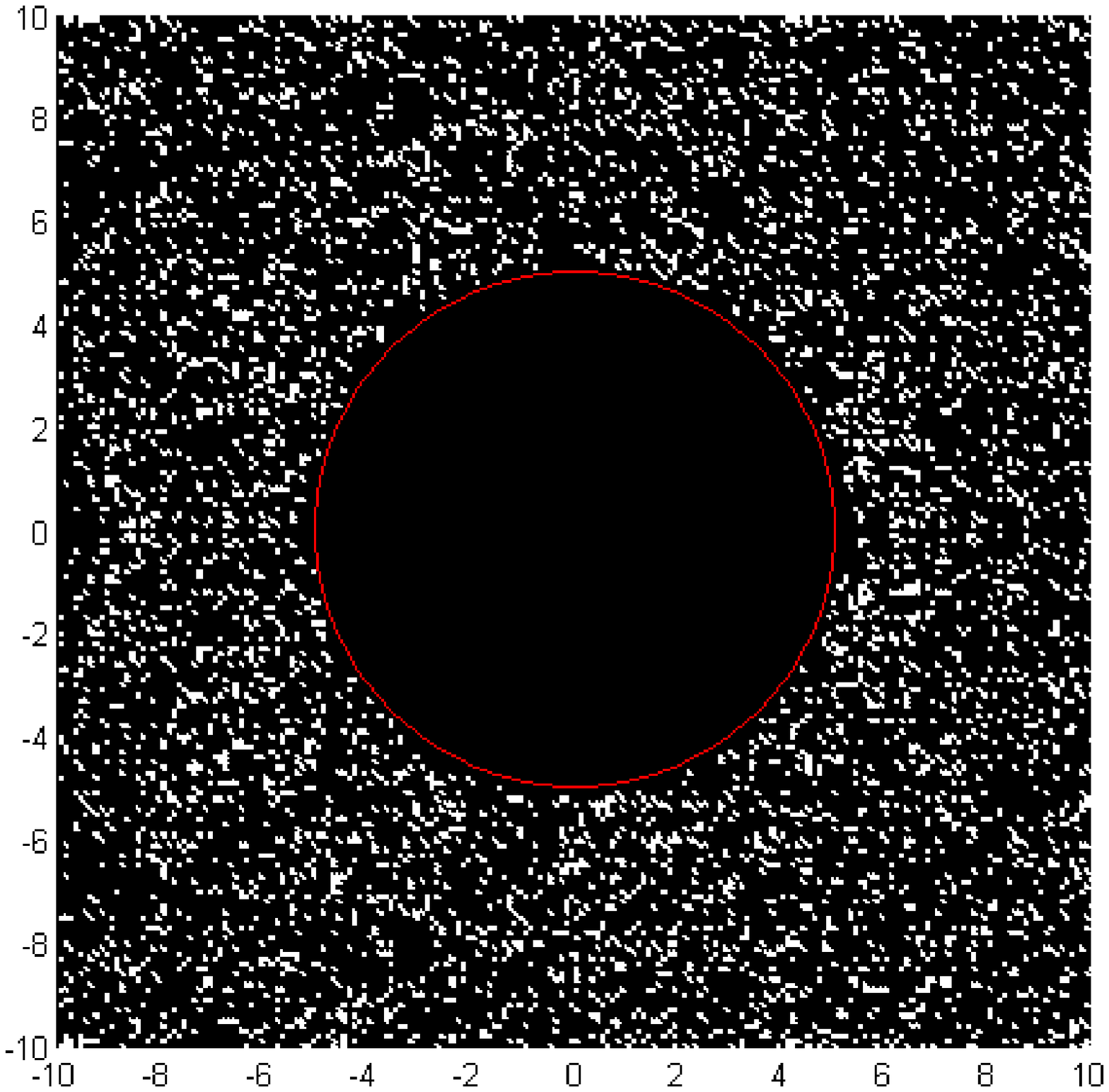} 
\epsfxsize=3.75cm \hspace{0.0cm} \epsfbox{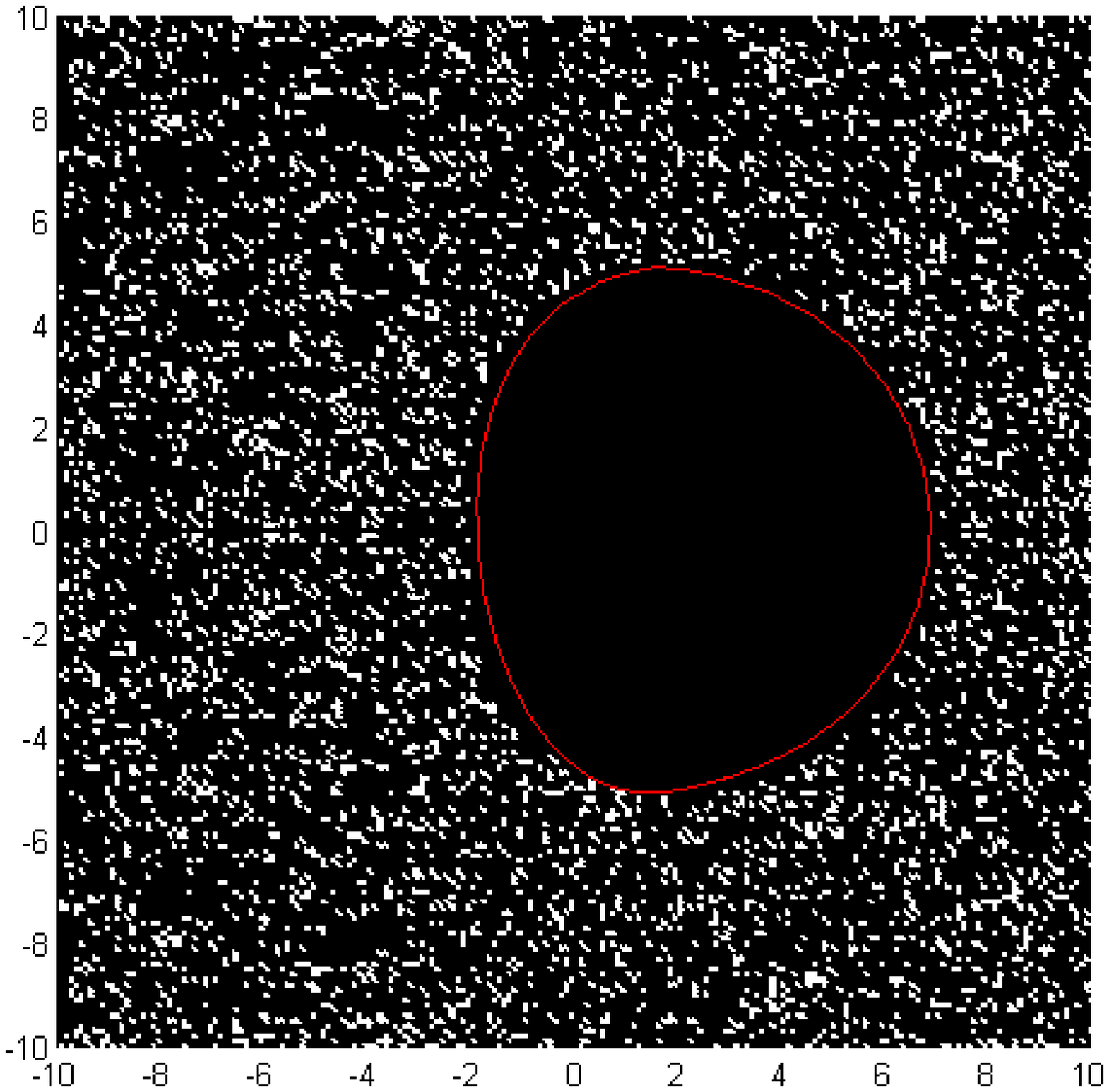}  
\epsfxsize=3.75cm \hspace{0.0cm} \epsfbox{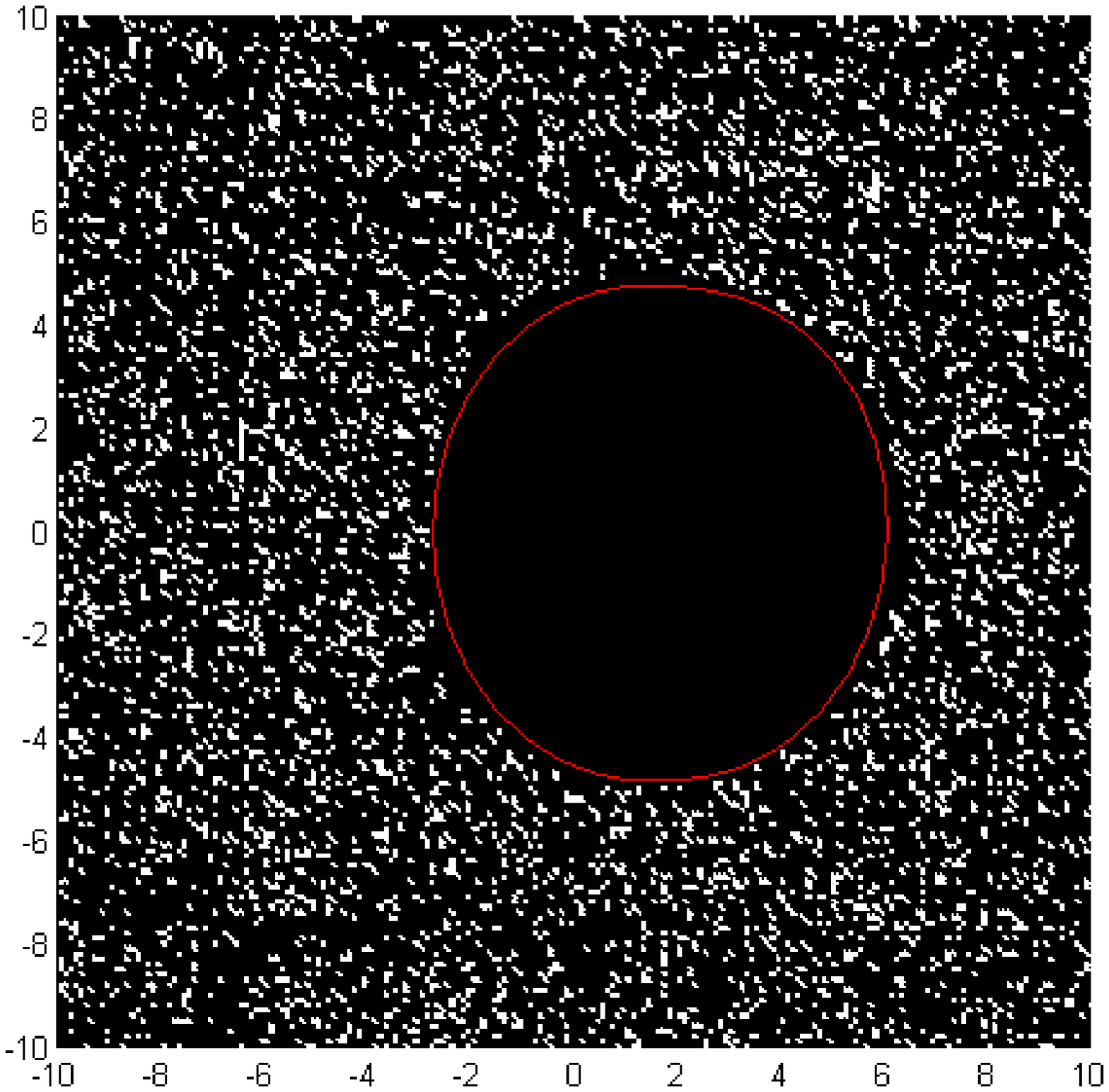} \\ 
\epsfxsize=3.75cm \hspace{0.0cm} \epsfbox{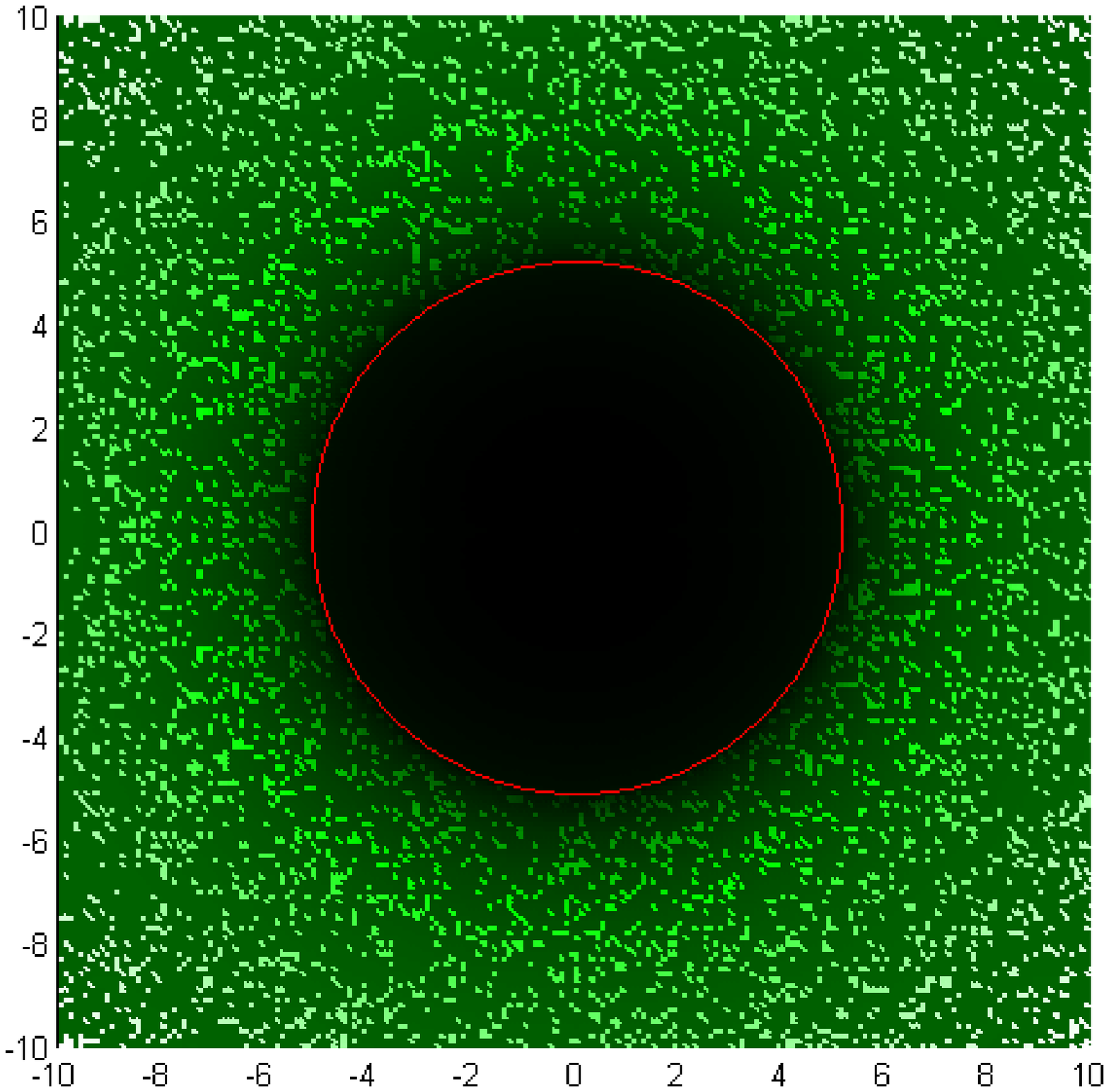}  
\epsfxsize=3.75cm \hspace{0.0cm} \epsfbox{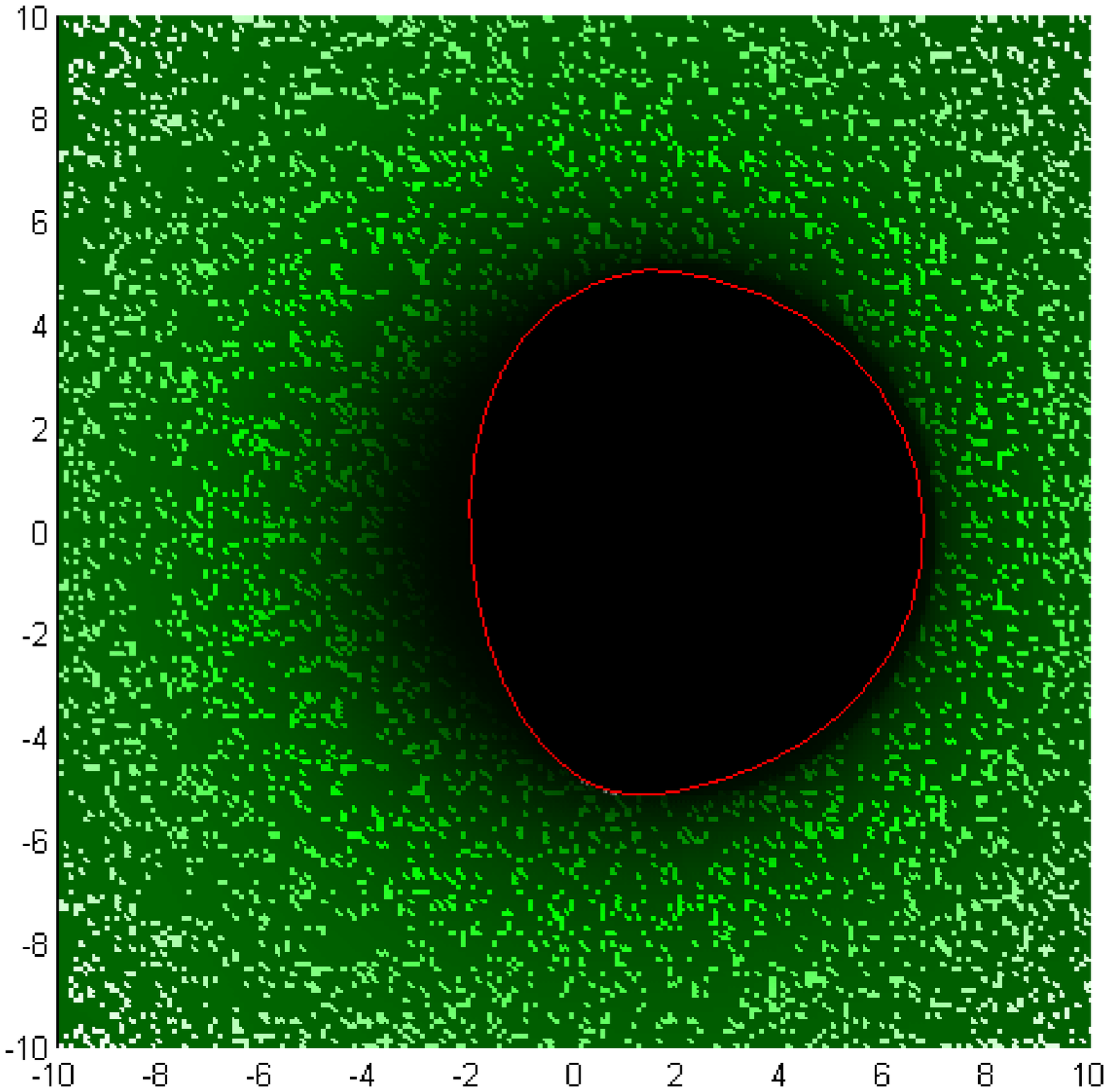} 
\epsfxsize=3.75cm \hspace{0.0cm} \epsfbox{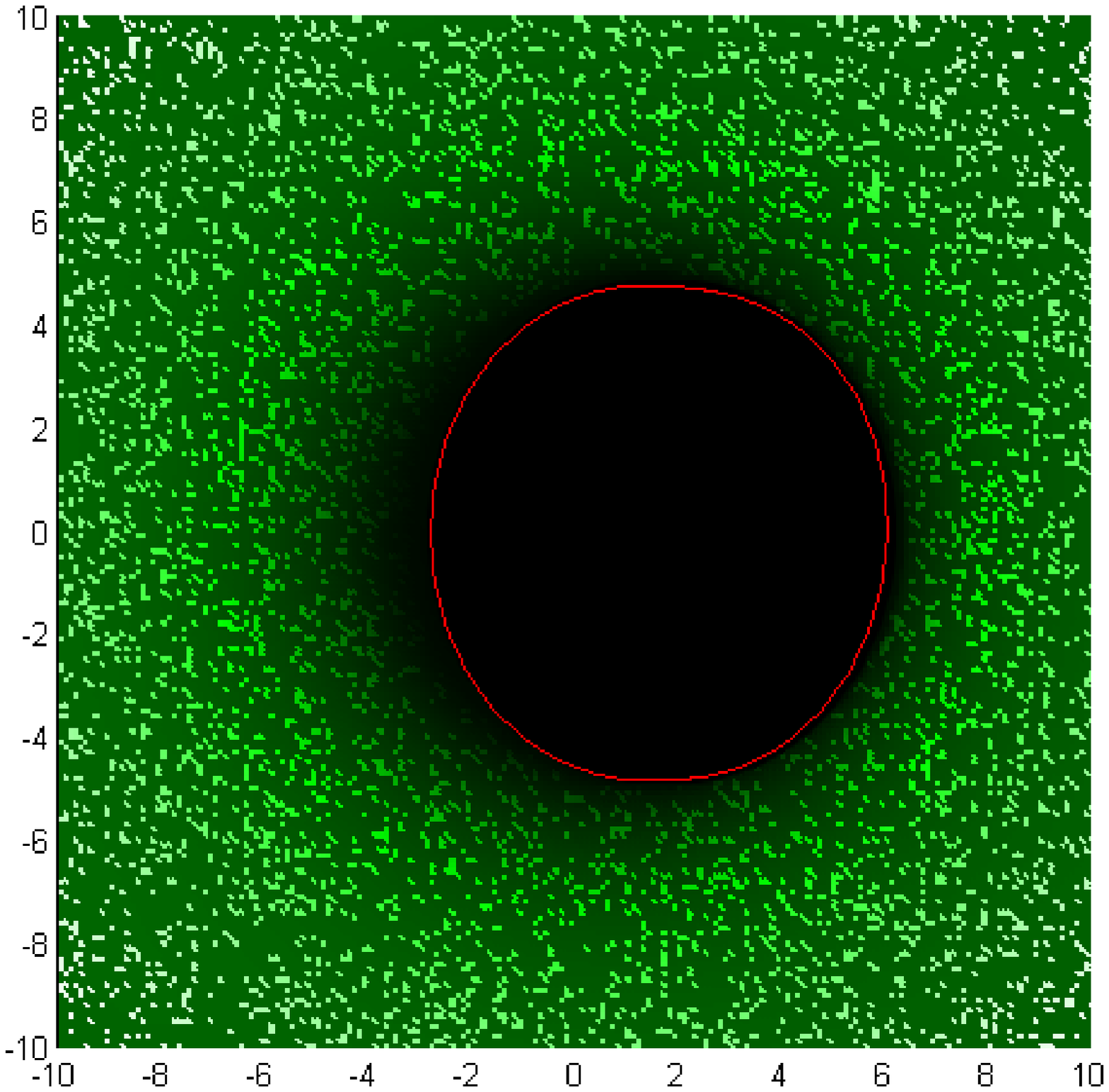}  
\end{center}
\vspace*{0cm} 
\caption[h]{Shadows of a background sky cast 
     by Schwarzschild black holes (left column),  
     and Kerr black holes with $a=0.998$ 
       viewed at inclination of $90^\circ$ (middle column) 
       and $45^\circ$ (right column).  
  The optical depth is provided by a spherical inflow of warm accreting gas 
    with a temperature $T = 5\times 10^5$~K.  
  The density distribution of the gas takes the form: $n(r) \propto r^{-3/2}$.        
  The line-of-sight opacity  $\tau = 0$ for the black holes in the top row,  
     and the mean line-of-sight opacity $\tau = 5$ for the black holes in the bottom row.  
  The stars in the background sky are randomly generated   
     and have the same brightness.  
  The red line denotes the locations of geodesics crossing the event horizon.  }
\end{figure}

\section{Imaging Black Holes}  

Figure 1 and 2 show the images of accretion tori around Kerr black holes.  
The opacities are provided by the Fe K$\alpha$ and K$\beta$ lines 
  and by electron scattering respectively in the two cases.   
For the case in Figure 1, the non-scattering formulation (Eq.~1) is used in the calculations; 
  for the case in Figure 2, the moment radiative transfer formulation  (Eq.~2, 3 and 4) is used. 
The velocity field and density distribution in the tori are determined 
  using a empirical velocity law gauged by magneto-rotational instability simulations 
  of accretion disks (see Fuerst and Wu 2004, 2007).     
The brightness distributions of the tori in the images reflect the combined effects 
  of gravitational and kinematic frequency shifts, Doppler boosting,       
  and line-of-sight path length variations caused 
  by gravitational lensing and frame dragging.   
  
Figure 3 shows the emission images of electron-positron annihilation lines 
  from hot accretion flows. 
We adopt a vertically Gaussian density distribution $n(r,z)  \propto z^{-3/2} \exp(-z^2/H^2)$, 
    where the scale height $H$ is determined by the thermal sound speed   
       and the rotational Keplerian angular velocity (see Ball 2008 for details).   
This density distribution is appropriate for hot radiative inefficient accretion flows.   
A similar density distribution was used in the black-hole shadowing calculations 
   by Huang et al.\ (2007).   
The emission images of the Schwarzschild and Kerr black holes are distinguishable 
  with the differences more obvious for larger viewing inclination angles.  

Figure 4 shows the shadows in a starry background sky 
  cast by black holes with a Bondi-Hoyle accretion flows, 
  i.e.\ the density distribution of material $n(r) \propto r^{-3/2}$.    
The emissions from the stars are gravitationally lensed, 
   but the stellar emissions have no frequency shifts because the stars are at infinite distances.   
However, the stars appear to distribute unevenly in the images 
  and the centroid of the shadow appears to shift when the black hole rotates. 
These phenomena happen for two reasons. 
The first is due to `geometrical effect', which is caused by rotational frame dragging of the black hole. 
This can be seen easily when comparing the images in the top panels. 
The shadow is not circular and the centroid shifts if the black hole rotates.  
  The distortion depends on both the black-hole spin and the viewing inclinations. 
The second is  due to `optical depth effect'. 
Gravitational lensing and frame dragging have different effects on the lines-of-sight 
    for different impact parameters and 
    for different photon incidence with respect to the black hole's spin axis 
    (in the case of Kerr black holes).  
The integrated optical depths are not the same along different lines-of-sight.  
In the bottom panels of Figure 4, 
  the emissions of the stars at the left side of the event horizon 
  are more strongly absorbed
  than the emissions of the stars at the right side.

\section*{Acknowledgements}
We thank Colin Young and Susan Young for reading through this manuscript and for comments.

\bigskip
\noindent {\bf DISCUSSION}

\bigskip
\noindent {\bf THOMAS SCHWEIZER:} Is it possible to measure the spin 
  of a black hole and the last stable orbit by looking at the emission lines?  

\bigskip
\noindent {\bf KINWAH WU:} In principle, yes. 
  In practice, it is more complicated to deduce the black-hole parameters  
      from the emission lines 
      because of degeneracy. 
  The emission line profiles  
      are not uniquely determined by the location of the last stable orbit of particles, 
      and hence the black hole's spin.   

\end{document}